\documentclass{article}

\PassOptionsToPackage{numbers, sort&compress}{natbib}

\usepackage[creativeai, final]{neurips_2025}

\usepackage[utf8]{inputenc} 
\usepackage[T1]{fontenc}    
\usepackage{hyperref}       
\usepackage{url}            
\usepackage{booktabs}       
\usepackage{amsfonts}       
\usepackage{nicefrac}       
\usepackage{microtype}      
\usepackage{xcolor}         
\usepackage{cleveref}

\usepackage{graphicx}

\newcommand{\Detailed}{Detailed}
\newcommand{\detailed}{detailed}

\title{Music Arena: Live Evaluation for Text-to-Music}

\author{%
  Yonghyun Kim$^{\sharp\sharp}$ \\
  \And
  Wayne Chi$^{\flat}$ \\
  \And
  Anastasios Angelopoulos$^{\natural}$ \\
  \And
  Wei-Lin Chiang$^{\natural}$ \\
  \AND
  Koichi Saito$^{\sharp}$ \\
  \And
  Shinji Watanabe$^{\flat}$ \\
  \And
  Yuki Mitsufuji$^{\sharp}$ \\
  \And
  Chris Donahue$^{\flat}$ \\
  \And
  \vspace{0.5em}
  \small
  \begin{tabular}{@{}c@{}}
    $^{\flat}$Carnegie Mellon University \quad
    $^{\natural}$LMArena \quad
    $^{\sharp}$Sony AI \quad
    $^{\sharp\sharp}$Georgia Tech \quad
  \end{tabular}
}

\begin{document}

\maketitle

\vspace{-8mm}
\begin{center}
    \textbf{Music Arena is available at: \url{https://music-arena.org}} \\
    \textbf{Preference data is available at: \url{https://huggingface.co/music-arena}}
\end{center}

\begin{abstract}

    We present Music Arena,
    an open platform for 
    scalable human preference evaluation of text-to-music (TTM) models. 
    Soliciting human preferences via listening studies is the gold standard for evaluation in TTM, 
    but these studies are expensive to conduct and 
    difficult to compare, as study protocols may differ across systems.  
    Moreover, human preferences might help researchers align their TTM systems or improve automatic evaluation metrics, 
    but an open and renewable source of preferences does not currently exist. 
    We aim to fill these gaps by offering \emph{live} evaluation for TTM. 
    In Music Arena, 
    real-world users input text prompts of their choosing and compare outputs from two TTM systems, 
    and their preferences are used to compile a leaderboard. 
    While Music Arena follows recent evaluation trends in other AI domains, we also design it with key features tailored to music:
    an LLM-based routing system to navigate the heterogeneous type signatures of TTM systems, 
    and the collection of \emph{\detailed} preferences including listening data and natural language feedback. 
    We also propose a rolling data release policy with user privacy guarantees, 
    providing a renewable source of preference data and increasing platform transparency.
    Through its standardized evaluation protocol, transparent data access policies, and music-specific features, 
    Music Arena not only addresses key challenges in the TTM ecosystem but also demonstrates how live evaluation can be thoughtfully adapted to unique characteristics of specific AI domains.
\end{abstract}
\section{Introduction}
\label{sec:intro}

Text-to-music (TTM) generation has advanced rapidly in recent years, with models demonstrating remarkable capabilities in creating high-fidelity music audio from text prompts~\citep{dhariwal2020jukebox,ForsgrenMartiros2022riffusion,agostinelli2023musiclm, copet2023simple, evans2025stable}. 
This progress has highlighted two critical and intertwined challenges for the research community. 
Firstly, 
designing \emph{rigorous TTM evaluation} protocols 
is essential for navigating tradeoffs in methodologies and training data, and also to track progress over time. 
Secondly, 
identifying a source of \emph{open and renewable human TTM preference data} would help researchers to better align TTM systems with human intent~\citep{cideron2024musicrl}, and 
aid in the development of more reliable automatic evaluation metrics~\citep{grotschla2025benchmarking,huang2025aligning}. 

The current TTM landscape is unable to meet these challenges. 
Music is a human endeavor, and a rigorous evaluation metric should thus reflect human preferences.
However, human preferences are difficult to capture as they may be influenced more by subjective assessments of creativity than objective, quantifiable phenomena. 
While numerous automatic evaluation metrics have been proposed~\citep{kilgour2019frechet, gui2024adapting, huang2025aligning}, 
past work shows they correlate imperfectly with human preferences~\citep{huang2025aligning,grotschla2025benchmarking} and may not capture all key musical desiderata~\citep{huang2025aligning}. 
Moreover, 
while some open preference datasets have been released~\citep{liu2025musiceval,huang2025aligning,grotschla2025benchmarking}, 
these one time efforts are not \emph{renewable} and will remain fixed even as new models emerge or human preferences drift. 
Commercial providers may have access to renewable sources of preferences through their platform's proprietary usage data, but this data is typically not \emph{open}.

Human listening studies could potentially address both gaps, 
offering evaluation grounded in human preferences and a source of preference data, 
but current listening protocols lack \emph{rigor}.  
Firstly, they are \emph{inconsistent}---the meaning of metrics like win rates or mean opinion scores varies across numerous dimensions of ad hoc protocols including listening interfaces, models compared, and annotator distributions. 
Secondly, studies routinely cost hundreds or thousands of dollars on crowdsourcing platforms, making them \emph{unscalable}. 
Finally, studies are \emph{unrealistic} as users are presented with contrived listening scenarios that differ from real-world, self-motivated usage of TTM systems.

A new evaluation protocol, which we refer to here as \emph{live evaluation}, has already helped navigate analogous challenges in other AI domains~\citep{chiang2024chatbot, tts-arena-v2, jiang2024genai, ebert20253d, chi2025copilot}. 
The key idea behind live evaluation is to align incentives by offering everyday users free access to generative AI systems in exchange for their preferences. 
On most live evaluation platforms, 
users first submit an input query, 
are presented with outputs from two different AI systems, and 
finally asked which of the two they prefer. 
These pairwise preferences are distilled into a global leaderboard ordered by Bradley-Terry coefficients~\citep{bradley1952rank} or related scores like Elo~\citep{elo1967proposed}. 
Compared to ad hoc human evaluation protocols, 
these scores are more \emph{consistent} because they are calculated for all models from the same protocol and annotator distribution. 
Moreover, collecting preferences in this fashion is \emph{scalable} because incentives are aligned, and 
the preference data reflects more \emph{realistic} usage. 
Live evaluation was first proposed for language via Chatbot Arena~\citep{chiang2024chatbot}, and subsequently explored for 
text-to-speech~\citep{tts-arena-v2}, 
image and video generation~\citep{jiang2024genai}, 
3D model generation~\citep{ebert20253d},
and coding assistance~\citep{chi2025copilot}. 
Here we propose to offer live evaluation for TTM.

\begin{figure}
    \centering
    \includegraphics[width=0.97\linewidth]{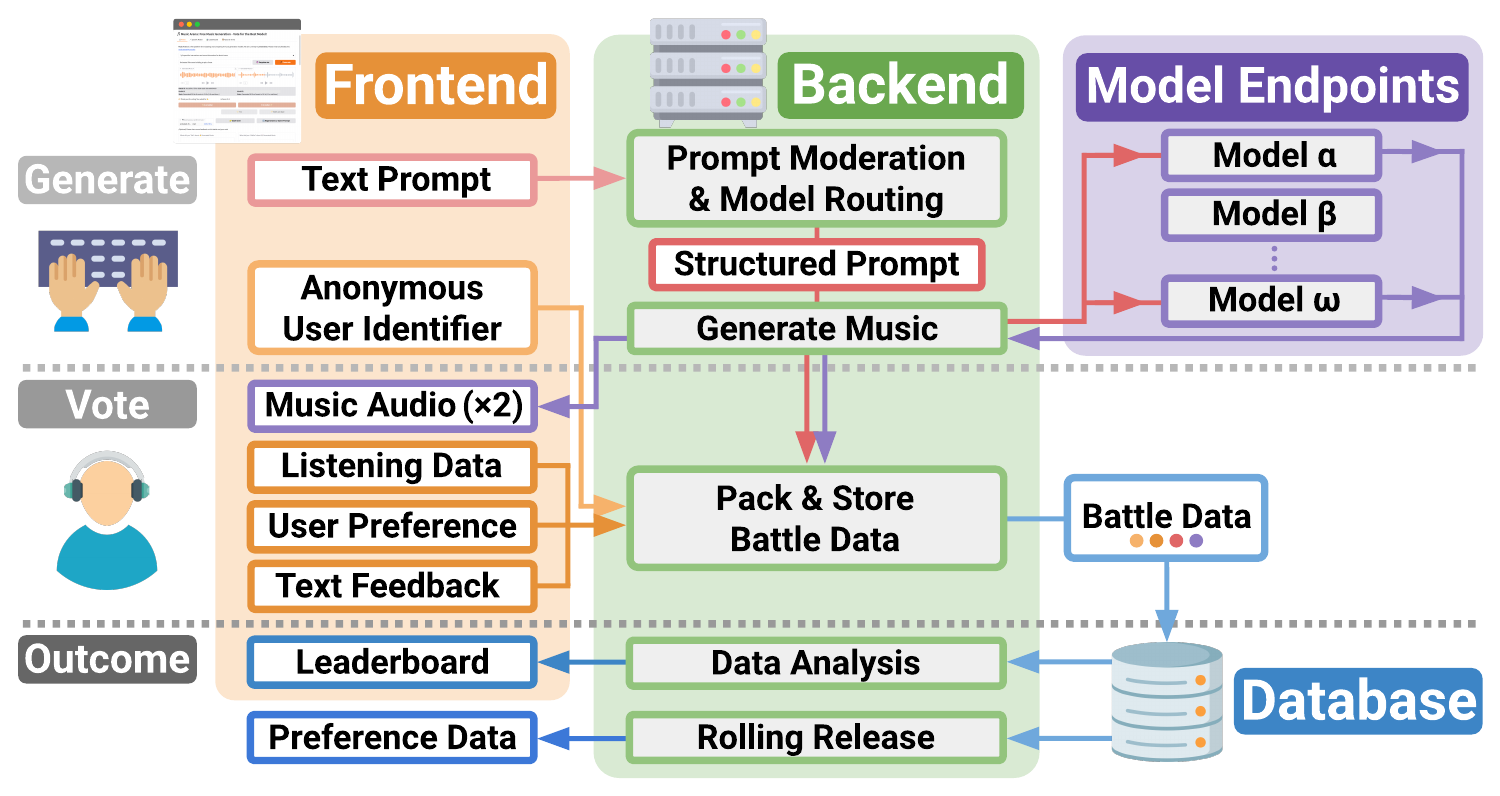}
    \caption{
    The Music Arena data lifecycle. 
    On the \textbf{Frontend}, users engage in ``battles'': they submit text prompts, listen to outputs from two music generation systems, and specify their preferences. 
    The central \textbf{Backend} orchestrates the battles: it 
    extracts structured information from text prompts using an LLM to determine model compatibility, 
    routes prompts to appropriate \textbf{Model Endpoints} for generation, delivers music audio to users, and 
    stores the resulting battle data in a \textbf{Database}. 
    Collected data is used to compile a public leaderboard and publicly released on a recurring basis.}
    \label{fig:ma-overview}
    \vspace{-4mm}
\end{figure}

Music presents unique challenges and opportunities for live evaluation relative to other AI domains. 
Firstly, music generation models have heterogeneous input and output type signatures: 
some models output vocals and may or may not accept user-specified lyrics, 
while other models output variable length audio and may or may not accept user-specified durations. 
We propose an LLM-based \emph{prompt routing} system which adapts inputs on a simple unified user interface (a single input text box) to appropriate type signatures for different models. 
We also use this system to detect and reject malicious inputs from users, including references to copyrighted material and artists, or overtly inappropriate or harmful content. 
Secondly, unlike for other modalities like images, music must be consumed by users in real time, affording the opportunity to collect data on which portions of an output a user observed before specifying their preferences. 
Accordingly, we collect \emph{fine-grained} listening data consisting of timestamps for playback actions on both outputs, and also enforce that users listen to some non-trivial amount of each before submitting their preferences. 
These key features of Music Arena underscore the importance of tailoring live evaluation to the nuances of individual domains.

We also implement policies for Music Arena that aim to both increase platform trust and provide renewable access to preference data.
Firstly, 
we anonymize private user information such as IP 
addresses by salting and hashing, 
ensuring \emph{user privacy} while also facilitating \emph{record linkage} for longitudinal preference research. 
Secondly, 
our entire platform is open source\footnote{Code is available at \url{https://github.com/gclef-cmu/music-arena}} aside from production configurations and secret keys. 
Finally, 
we propose to release data at \emph{regular} (monthly) intervals, 
allowing the research community to access the latest data. 
By open sourcing our code and data, the entire lifecycle of our platform and leaderboard can be independently audited.


We release Music Arena to address key challenges in the landscape of TTM (rigorous evaluation, renewable preference data). Here, we present an overview of the Music Arena platform alongside an analysis of the preferences collected between July 28 and Aug 31, 2025, including 1,420 user-initiated battles from 373 unique users yielding 1,051 votes. From these votes, we build a preliminary \textbf{Music Arena leaderboard (\Cref{sec:leaderboard})}, ranking contemporary TTM models via live evaluation.
\section{Music Arena Platform Overview}
\label{sec:overview}

Music Arena is a web-based live evaluation platform designed for the scalable collection of pairwise human preferences for TTM systems. 
The platform's architecture is comprised of three core components (\Cref{fig:ma-overview}): 
a user-facing \textbf{Frontend}, 
an intelligent \textbf{Backend} that orchestrates the entire generation and logging process, 
and \textbf{Model Endpoints} comprising the various TTM systems. 
These components are modular and communicate with one another via simple HTTP requests. 

\subsection{Frontend}
\label{sec:frontend}

Our frontend is a web-based 
interface built with Gradio~\citep{abid2019gradio} and serves as the primary means of user interaction. 
Upon their first visit, users are presented with a consent page detailing the IRB-approved research protocol and data handling policies. Once consent is provided, 
the main ``Arena'' interface allows users to engage in one or more ``Battles'' (pairwise comparisons). 
To initiate a battle, a user submits a text prompt of their choosing. 
Next, the user is presented with two audio tracks generated by two different TTM systems, and may listen to them in whatever order and for however long they like. 
To mitigate potential biases that could arise from differing audio lengths, the interface supports variable-length outputs but conceals the specific duration of each track from the user. 
After listening, 
users specify their preference (prefer A, prefer B, tie, both bad). 
Finally, after voting, the frontend 
reveals the identities of the competing models, along with other information such as generation speed.
A download link for the preferred track is provided as an incentive for casting a decisive vote, and users are given the option to provide additional natural language feedback. 

\subsection{Backend}
\label{sec:backend}

The backend is the central server-side component that acts as the main orchestrator for the platform. 
It receives all incoming requests from the frontend and is designed to handle numerous user sessions concurrently. 
Its core responsibility is to manage the entire lifecycle of each battle: 
it processes the user's text prompt and dispatches generation tasks to two models in parallel. 
To mitigate bias from differing inference speeds, the backend waits for both models to complete and then delivers their audio tracks simultaneously---the actual generation time of each model is also logged on the backend. 
Finally, it collects the user's preference data and ensures it is securely stored in the database.

\subsection{Model Endpoints} 
\label{sec:endpoints}

To facilitate Music Arena, we aim to unify the heterogeneous type signatures of TTM systems. 
To this end, we implement a \emph{model endpoint} for numerous TTM systems~\citep{copet2023simple,evans2025stable,liu2025songgen,gong2025ace,google2025magenta_rt,google2025lyria_rt,ForsgrenMartiros2022riffusion}---code that adapts underlying type signatures and dependencies into a common interface. 
For open weights models, endpoints manage synchronous calls to GPU resources and batching for increased throughput. 
For API-based commercial systems, endpoints adapt inputs from our unified API to HTTP requests to proprietary APIs. 
To manage the varied (and sometimes conflicting) software dependencies across systems, 
each endpoint is paired with a bespoke Docker container. 
Each container exposes a simple API endpoint with a common type signature, 
allowing the backend to interact with all systems in a uniform manner. 
This modular architecture facilitates decentralized development---providers can contribute new TTM systems to Music Arena without disturbing other parts of the platform. 
\section{Key Features}
\label{sec:features}

Here we detail the methods and policies we propose in Music Arena, especially those distinct from other live evaluation platforms or ones that are specifically tailored for the music domain.

\subsection{LLM-based moderation and routing}
\label{sec:llm}

The backend's orchestration is powered by an LLM-based system that facilitates \emph{moderation} of malicious input prompts, and 
.0\emph{routing} of prompts on a unified interface (simple textbox) to relevant models. 
To moderate, the LLM is instructed to reject the user's input prompt if it contains references to 
copyrighted musical material, 
culturally insensitive themes, 
or explicit themes, including profanity that would be atypical for the musical style 
(e.g.,~profanity okay for heavy metal, not okay for a nursery rhyme). 
For prompts that pass moderation, 
the LLM is instructed to extract structured information from the natural language input: 
the implied presence of vocals or lyrics (e.g.,~``folk song about a cat named Chamomile'' implies lyrics), and
explicitly-specified duration (e.g.,~``$30$ second lo-fi beat''). 
The backend uses this structured representation to seamlessly route prompts to a subset of models that support the user's query (e.g.,~many models do not support vocal or lyrics generation). 
At time of writing, Music Arena 
uses OpenAI's GPT-4o~\citep{hurst2024gpt} for this component. 

\subsection{\Detailed{} preferences via listening data and language feedback}
\label{sec:fgprefs}

Most live evaluation platforms for other AI domains collect simple pairwise preferences. 
Here we propose to additionally collect more detailed preference signals including fine-grained listening data and natural language feedback. 
As the user listens to each generated audio during a battle, 
our system stores their listening behavior including the amount of time spent listening to each clip, and the wall clock time at which they played or paused each clip. 
To ensure meaningful user engagement, 
the voting interface is only unlocked after a user listens to each track for a predefined minimum duration ($4$ seconds at the time of writing). 
After a user specifies their preference between four options---\emph{A is better}, \emph{B is better}, \emph{Tie}, or \emph{Both are bad}---they are encouraged to provide additional natural language feedback, clarifying their rationale. 
We
use listening data to better understand and model user behavior, and the language feedback to offer richer insights into preferences and desiderata than binary preferences alone can provide. 
An example of our detailed preference data per battle and an analysis of the initial data release 
appear in ~\Cref{sec:battle_example} and ~\Cref{sec:analysis}, respectively.

\subsection{Reference TTM implementations}
\label{sec:models}

A key feature of Music Arena is the development of a unified Docker-based framework for managing inference from TTM systems as outlined in~\Cref{sec:endpoints}. 
In addition to supporting our core platform, 
we hope that this unified framework may benefit other research that requires comparing outputs from several TTM systems. 
At time of writing, we support the following open weights models: 
Meta's MusicGen~\citep{copet2023simple}, 
Stability AI's Stable Audio Open~\citep{evans2025stable} and Stable Audio Open Small~\citep{novack2025fast}, 
SongGen~\citep{liu2025songgen}, 
ACE Studio's ACE-Step~\citep{gong2025ace}, and
Google DeepMind's Magenta RealTime~\citep{google2025magenta_rt}. 
We also support API-based commercial models including 
Producer.ai's FUZZ models (1.0 \& 1.1)~\citep{riffusion2025fuzz},
Stability AI's Stable Audio 2.0~\citep{evans2024long}, and
Google DeepMind's Lyria RealTime~\citep{google2025lyria_rt}. 
Due to resource limitations, 
not all of these models will be available for live evaluation at a given time, 
however they can always be accessed by researchers running our code using their own resources.

Systems in this collection exhibit substantial heterogeneity in type signatures. 
Three support generating output vocals~\citep{liu2025songgen,gong2025ace,riffusion2025fuzz} while others are instrumental only. 
Commercial systems like Riffusion FUZZ~\citep{riffusion2025fuzz} generate lyrics jointly with audio, 
while open weights systems like SongGen~\citep{liu2025songgen} and ACE-Step~\citep{gong2025ace} require explicit lyrics conditioning---
we use GPT-4o~\citep{hurst2024gpt} to generate lyrics for these systems from a user's input prompt. 
In addition to considerations around vocals, there is a long tail of additional control signals across models, e.g.,~Stability AI models support explicit specification of output duration~\citep{evans2025stable,novack2025fast,evans2024long}. 
This heterogeneity is more pronounced in music, unlike the standardized signatures of chat~\citep{chiang2024chatbot} and image generation~\citep{jiang2024genai}.
We design our unified framework to navigate this complex landscape, 
and aim to extend it in future work to support even broader music type signatures beyond TTM such as style transfer or symbolic music generation.
\section{Key Policies}
\label{sec:policies}

Along with the key \emph{features} outlined previously, 
here we emphasize key \emph{policies} of Music Arena designed to increase platform trust by 
surfacing considerations beyond user preferences, 
promoting user privacy and platform transparency, and 
providing renewable access to preference data.

\subsection{Surfacing model considerations beyond Arena Score}
The primary metric of interest across existing live evaluation platforms in other domains is an ``Arena Score'', derived from pairwise comparisons. 
In music, holistic comparisons may require considerations of factors beyond preferences. 
Accordingly, our leaderboard (\Cref{sec:leaderboard}) also surfaces information on \textbf{training data} and \textbf{generation speed}. 
Including information on training data acknowledges the legal, ethical, and quality implications of the ``uneven playing field'' for training data across model providers. Reporting generation speed (measured by median RTF) recognizes that some models may trade off quality for speed, e.g., to facilitate low-latency creative workflows.

\subsection{Protecting user privacy while facilitating record linkage}
\label{sec:privacy}

The ability to perform \emph{record linkage} across Music Arena sessions---identifying multiple battles from the same user---is critical for longitudinal preference analysis and detecting spam or malicious behavior.
However, it is essential that we also protect user privacy by ensuring that personally identifying information is never exposed.
Following established recommendations for user privacy in research~\citep{tene2012privacy}, 
we implement a pseudonymization protocol using salted hashing~\citep{kushida2012strategies}. 
When users interact with Music Arena, we transform linkable identifiers such as IP addresses by applying a server-side salt (a secret random string) followed by a one-way cryptographic hash. 
We only store these anonymized user identifiers, never the original identifiers. 
This approach provides strong privacy guarantees and protects against de-anonymization strategies such as rainbow table attacks. 
The resulting anonymized identifiers allow anyone to link battles from the same user across sessions for research purposes without ever exposing their private identifiers.

\subsection{Maximizing platform transparency and data access}
\label{sec:transparency}

We are committed to making Music Arena as \emph{transparent} as possible. 
To this end, 
all of our platform code is \textbf{open source}, 
aside from secret keys for private salting and API access. 
Additionally, 
we commit to a policy of \textbf{rolling, comprehensive data releases}. 
Unlike preferences from one-time data collection efforts~\citep{liu2025musiceval,huang2025aligning,grotschla2025benchmarking},
we aim to publish Music Arena data at regular monthly intervals.
This rolling approach is critical in the rapidly evolving field of generative AI,
addressing key sources of distribution shift like 
the development of new TTM systems, and
changes in user preferences over time. 
Moreover, 
we aim for our data releases to be comprehensive,\footnote{Minor exceptions to this policy may apply, e.g.,~some model licenses prevent the release of generated audio.} including anonymized user identifiers, generated audio, and \detailed{} preferences.
This combination of open code and data allows the research community to audit our platform's entire lifecycle and evaluation results.
\section{Ethical Considerations and Safeguards}
\label{sec:ethics}

The design and operation of Music Arena are guided by principles of ethical research, user privacy, and responsible AI development. All research activities involving human subjects in this study were approved by the IRB at Carnegie Mellon University under Protocol ID \texttt{STUDY2024\_00000489}. 
We have implemented several key safeguards in Music Arena to uphold these principles.

\textbf{Informed consent.} Before any interaction, users are presented with a consent page that transparently outlines the study's objectives, data collection methods, and our commitment to public data release. 
Explicit, informed consent is required to participate.

\textbf{User privacy.} Music Arena does not store personally identifiable information, such as raw IP addresses. Instead, we collect anonymized identifiers through salting and hashing. Additionally, users consent that they will not upload private information in their text prompts or language feedback.

\textbf{Content moderation.} 
To mitigate risks of harmful or infringing content creation through Music Arena, 
every user-submitted prompt is first processed by our LLM-based moderation pipeline.


There remain ethical considerations for our work beyond these safeguards. 
Music Arena provides increased access to TTM systems for everyday users, 
which could have long term psychological effects. 
Our user distribution will likely be skewed to US users and AI enthusiasts, 
potentially promoting increased focus to the needs of those user populations by model providers. 
Music Arena's current focus on text-to-music may inadvertently steer the research community's attention away from other important tasks like style transfer or symbolic generation.
Music Arena also inherits many ethical and societal considerations from music generation more broadly. 
Music generation may 
change the economic landscape of music labor, 
accelerate the commodification of music, or
contribute to the homogenization of music cultures. 
Overall, we believe the benefits of Music Arena (more rigorous and transparent evaluation, open availability of preference data) outweigh the risks. 

\section{Limitations and Future Work}
\label{sec:limitations}

Our work has several limitations that present clear avenues for future improvement. On the frontend, 
our ability to precisely track which segment of the audio a user is listening to is constrained by the user interface---we track total listening duration but cannot capture seeking actions within audio clips. 
Our current backend system also selects pairs of TTM systems uniformly at random, 
rather than more principled strategies~\citep{chi2025copilot} that navigate tradeoffs around quality, speed, and coverage of relevant pairs. 
Furthermore, the scope of Music Arena is currently limited to text-to-music generation, excluding other important tasks like symbolic generation or style transfer. 
Finally,
as a public web platform, 
our user base is not representative of the global population, and the long-term sustainability of providing free access to self-hosted open-weights models remains a challenge.

Our plans for future work aim to address these limitations, improve our understanding of human musical preferences, and contribute to the science of live evaluation.
We aim to continue refining our frontend and backend---a particular direction of interest is improving the backend pair selection algorithm to better balance tradeoffs around leaderboard fidelity and user experience. 
We will leverage the growing preference dataset to better understand strengths and weaknesses of specific models and perform meta-evaluation against automatic evaluation metrics. 
Through analysis of natural language feedback and live evaluation of controlled degradations of systems (e.g., adding latency or noise to a system), 
we may better understand which attributes users consider most prominently when making preference decisions. 
As creative workflows mature,
we hope to integrate live evaluation directly into user workflows~\citep{chi2025copilot}. 
Finally, we will continuously refine our evaluation methodology based on community feedback to ensure the long-term rigor and fairness of our platform.
\section{Conclusion}

We present Music Arena, a live evaluation platform that addresses critical gaps in text-to-music evaluation through scalable human preference collection and transparent data releases. 
Our platform introduces key innovations tailored specifically for music: 
an LLM-based system that enables content moderation and intelligent routing across heterogeneous model type signatures, 
a \detailed{} preference collection methodology that captures fine-grained listening behaviors and natural language feedback, and a commitment to open science through comprehensive rolling data releases and full lifecycle auditability. 
By aligning user incentives with research needs, Music Arena provides the community with both a standardized human evaluation protocol and a renewable dataset of human musical preferences that reflects real-world usage patterns. 
As text-to-music generation continues to advance rapidly, 
Music Arena establishes a foundation for rigorous evaluation that can evolve alongside the field, 
supporting researchers in building more aligned systems while maintaining transparency and ethical standards that respect both user privacy and the broader implications of AI-generated music.
\begin{ack}
The development of Music Arena was supported by Sony AI. 
We extend our sincere thanks to our commercial contacts at 
Producer.ai, Stability AI, Google DeepMind, and Suno for productive discussions that informed the key features and policies of Music Arena.
\end{ack}

\bibliographystyle{unsrtnat}
\bibliography{references}

\clearpage

\appendix

\section{Leaderboard structure}
\label{sec:leaderboard}

\begin{table}[t]
\centering
\caption{The first Music Arena leaderboard results (July 28 -- Aug 31, 2025), segmented by Instrumental and Vocal models. The table includes Arena Scores (with 95\% confidence intervals), vote counts, generation speed (RTF), and key metadata for each model.}
\label{tab:leaderboard_results}

\textbf{Instrumental Music Generation}
\vspace{2mm}

\resizebox{\textwidth}{!}{
\begin{tabular}{c c r c c c c c c c c}
\hline
\textbf{Rank} & \textbf{Model} & \textbf{\shortstack{Arena \\ Score}} & \textbf{95\% CI} & \textbf{\# Votes} & \textbf{\shortstack{Generation \\ Speed (RTF)}} & \textbf{Organization} & \textbf{License} & \textbf{\shortstack{Training \\ Data}} & \textbf{\shortstack{Supports \\ Lyrics}} & \textbf{Access} \\
\hline
1 & riffusion-fuzz-1-1 & 1250.8 & \shortstack{+52.0 / \\ -45.5} & 252 & 6.01 & Producer.ai & Closed & Unspecified & True & Proprietary \\
2 & magenta-rt-large & 1113.6 & \shortstack{+56.5 / \\ -57.2} & 276 & 1.01 & Google DeepMind & Apache 2.0 & Stock & False & Open weights \\
3 & musicgen-small & 928.5 & \shortstack{+40.4 / \\ -46.7} & 278 & 0.86 & Meta & CC-BY-NC 4.0 & Stock & False & Open weights \\
4 & sao & 924.7 & \shortstack{+45.7 / \\ -41.5} & 286 & 2.63 & Stability AI & STAI Community & Open & False & Open weights \\
5 & sao-small & 782.4 & \shortstack{+50.9 / \\ -62.2} & 292 & 12.79 & Stability AI & STAI Community & Open & False & Open weights \\
\hline
\end{tabular}
}

\vspace{5mm} 

\textbf{Vocal Music Generation}
\vspace{2mm}

\resizebox{\textwidth}{!}{
\begin{tabular}{c c c c c c c c c c c}
\hline
\textbf{Rank} & \textbf{Model} & \textbf{\shortstack{Arena \\ Score}} & \textbf{95\% CI} & \textbf{\# Votes} & \textbf{\shortstack{Generation \\ Speed (RTF)}} & \textbf{Organization} & \textbf{License} & \textbf{\shortstack{Training \\ Data}} & \textbf{\shortstack{Supports \\ Lyrics}} & \textbf{Access} \\
\hline
1 & riffusion-fuzz-1-0 & 1172.5 & \shortstack{+99.1 / \\ -62.7} & 144 & 5.60 & Producer.ai & Closed & Unspecified & True & Proprietary \\
2 & riffusion-fuzz-1-1 & 1087.3 & \shortstack{+40.8 / \\ -47.2} & 218 & 5.25 & Producer.ai & Closed & Unspecified & True & Proprietary \\
3 & preview-ocelot & 1045.7 & \shortstack{+75.9 / \\ -82.9} & 90 & 5.42 & Hidden & Closed & Unspecified & True & Proprietary \\
4 & preview-jerboa & 1034.4 & \shortstack{+92.6 / \\ -80.8} & 88 & 5.61 & Hidden & Closed & Unspecified & True & Proprietary \\
5 & acestep & 660.1 & \shortstack{+75.5 / \\ -121.3} & 178 & 2.89 & ACE Studio & Apache 2.0 & Unspecified & True & Open weights \\
\hline
\end{tabular}
}

\end{table}

\begin{figure}[t]
    \centering
    \includegraphics[width=\textwidth]{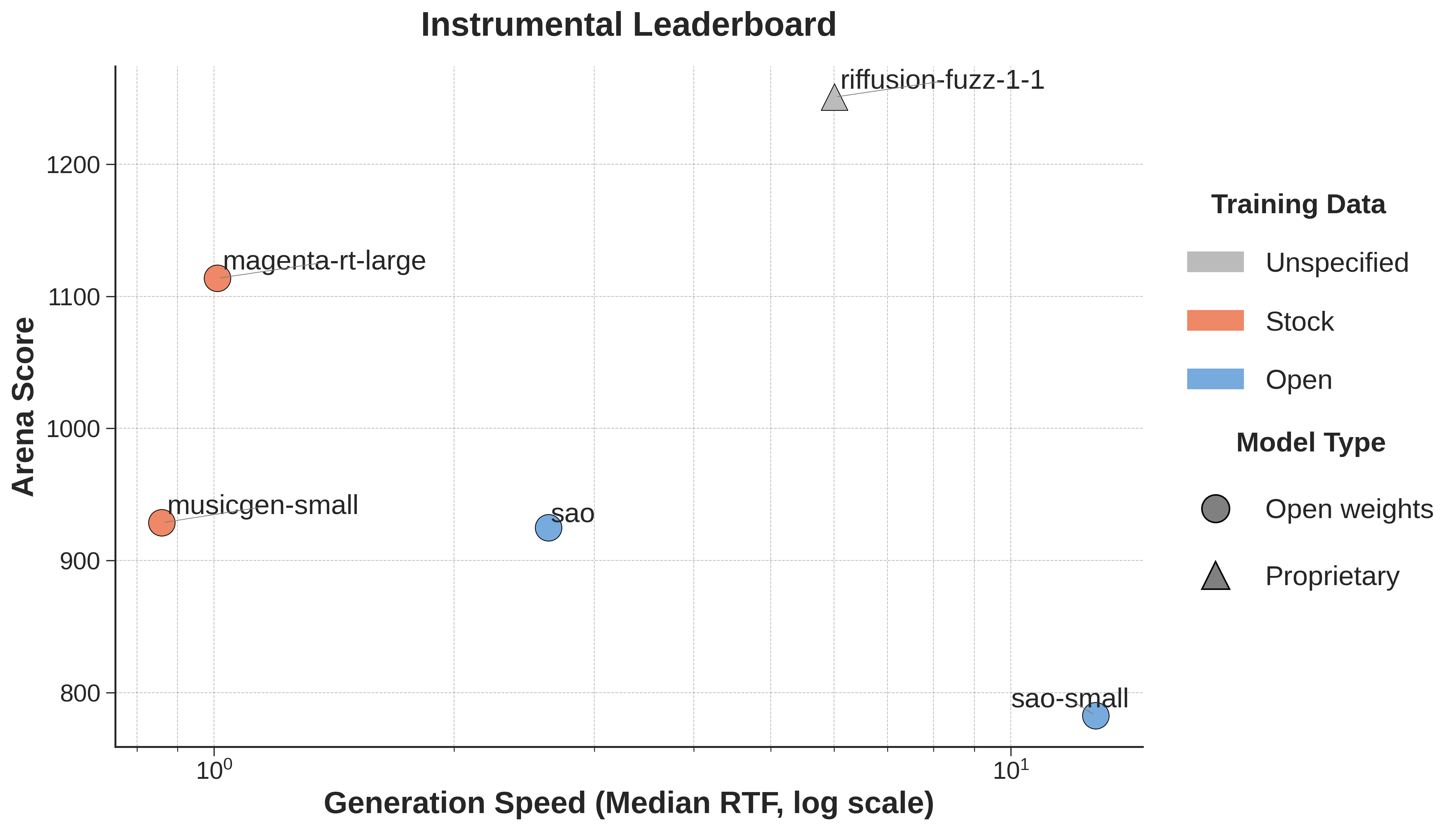}
    \includegraphics[width=\textwidth]{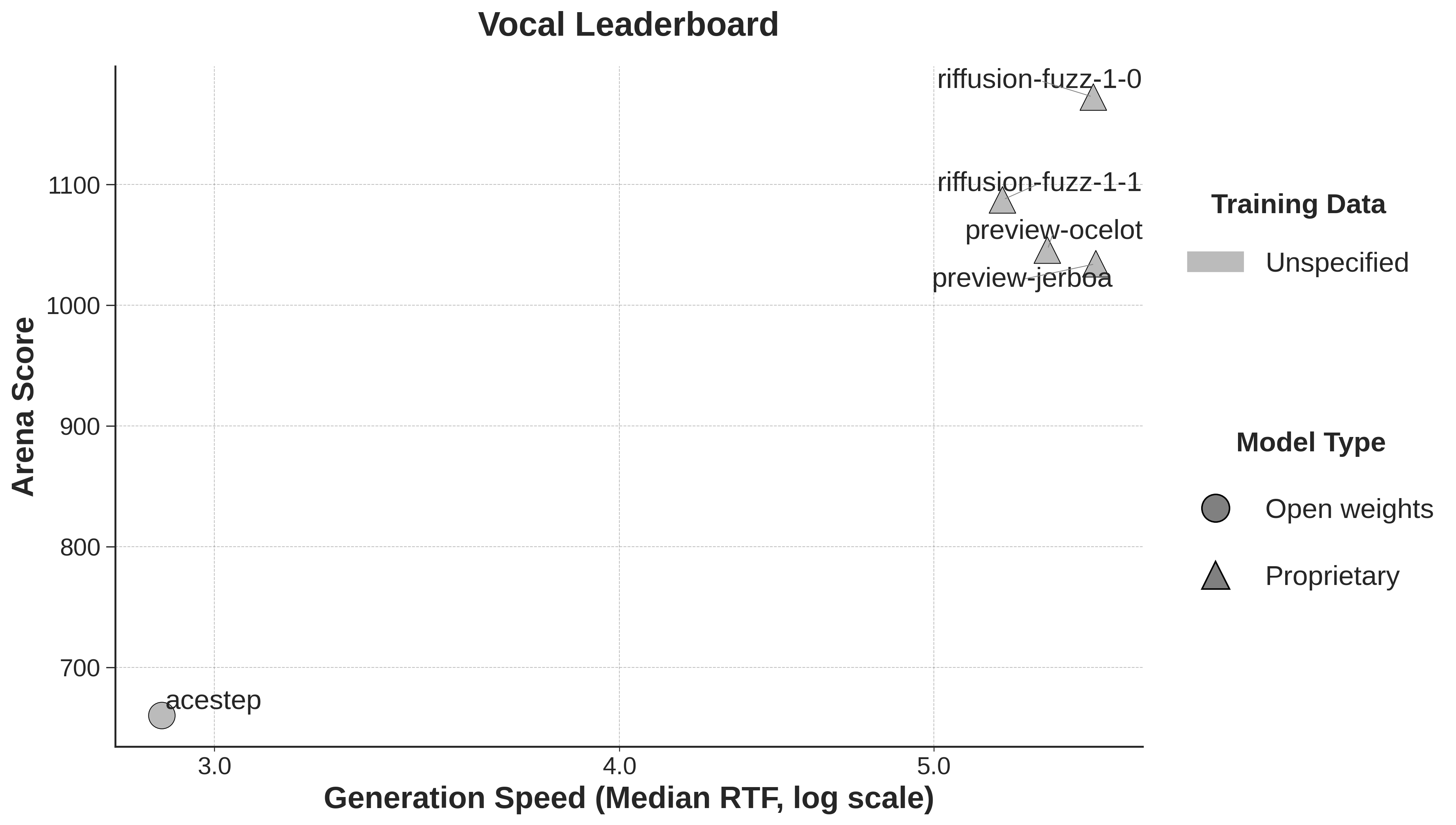}
    \caption{Music Arena leaderboard (July 28 -- Aug 31, 2025), plotting Arena Score (Y-axis) against Generation Speed (Median RTF, X-axis, log scale). Colors and shapes distinguish models by their training data and access type (open weights/proprietary), respectively. This visualization emphasizes the key tradeoff between model quality (score) and interactive latency (speed), an important consideration for creative music applications.}
    \label{fig:leaderboard}
\end{figure}



Music Arena was launched on July 28, 2025, and after collecting 1,051 user votes by the end of August, 
we released the first public leaderboard on September 19, 2025. Here, we provide an overview of the leaderboard's structure, which is designed to address the evaluation challenges in the music domain. The detailed results of this initial leaderboard are presented in \Cref{tab:leaderboard_results}, and the key tradeoffs are visualized in \Cref{fig:leaderboard}.

Public leaderboards for live evaluation platforms~\citep{chiang2024chatbot,jiang2024genai,chi2025copilot,tts-arena-v2} tend to contain similar attributes: 
an overall ``Arena Score'' (usually derived from the Bradley-Terry model~\citep{bradley1952rank}), 
the number of votes, 
the model provider, and the 
system license. 
There are a number of unique considerations in music that motivate presentation of additional attributes. 

For LLM training, 
it can be broadly assumed that all providers are training on large-scale text data mined from the web. 
However, in music, 
we see significantly more diversity in training data across models. 
For example, some models are trained on licensed stock music~\citep{copet2023simple} or publicly-available music under Creative Commons licenses~\citep{evans2025stable}, while some only specify the quantity (rather than the provenance) of their training data~\citep{gong2025ace}. 
These differences induce an uneven playing field for training data, 
affecting not only a system's performance in pairwise comparisons but also its standing within broader legal and ethical conversations. 
Accordingly, 
we include
on our leaderboard a summary of available \emph{training data information} for each model, including provenance and quantity.

In addition to training data, we 
include \emph{generation speed} on our leaderboard. 
Tools that facilitate the creation of music are often designed to have low latency interactions. 
Accordingly, TTM providers may make different tradeoffs between quality and speed, 
depending on if they are targeting more consumption-oriented (prefer quality) or creative-oriented (prefer speed) applications. 
Speed 
is codified on our leaderboard by median real-time factor (RTF), where RTF is the ratio of seconds of music generated divided by seconds of wall clock time to generate. 
For example, a system that generates $30$ seconds of audio in $3$ seconds has an RTF of $10$x. 
Measuring speed via RTF gracefully handles variable length outputs, ensuring that systems are not punished for taking more time to generate longer audio.

The leaderboard interface allows users to view results segmented by model capability (e.g.,~``Instrumental'' vs. ``Vocal'' tabs),
allowing viewers to make principled decisions about models based on attributes of keen importance to their specific music application goals. 
Moreover, 
we emphasize
these music-specific tradeoffs through visualization: 
a $2$D scatter plot with speed on the X axis, Arena Score on the Y axis, and colors and shapes to emphasize training data and licensing information. 

\section{Analysis of Initial Data Release}
\label{sec:analysis}
This section provides a more detailed look into user engagement patterns, prompt characteristics, and the types of music users create on the Music Arena platform.

\subsection{User Engagement Distribution} Our analysis shows a long-tail distribution of user engagement. While many users contribute a small number of votes, a dedicated group of ``power users'' is responsible for a significant portion of the data. \Cref{tab:user-engagement} shows the distribution of votes submitted per user from our initial data release.

\begin{table}[htbp]
\centering
\caption{Distribution of votes per user, based on 1,051 valid votes from 373 unique users.}
\label{tab:user-engagement}
\begin{tabular}{cc}
\hline
\textbf{Number of Votes} & \textbf{Number of Users} \\ \hline
1 & 193 \\
2 & 72 \\
3 & 44 \\
4 & 24 \\
5 & 8 \\
6-10 & 18 \\
11-20 & 10 \\
21-50 & 4 \\ \hline
\end{tabular}
\end{table}

\subsection{Prompt Descriptiveness} 
\begin{figure}[htbp]
    \centering    \includegraphics[width=0.8\linewidth]{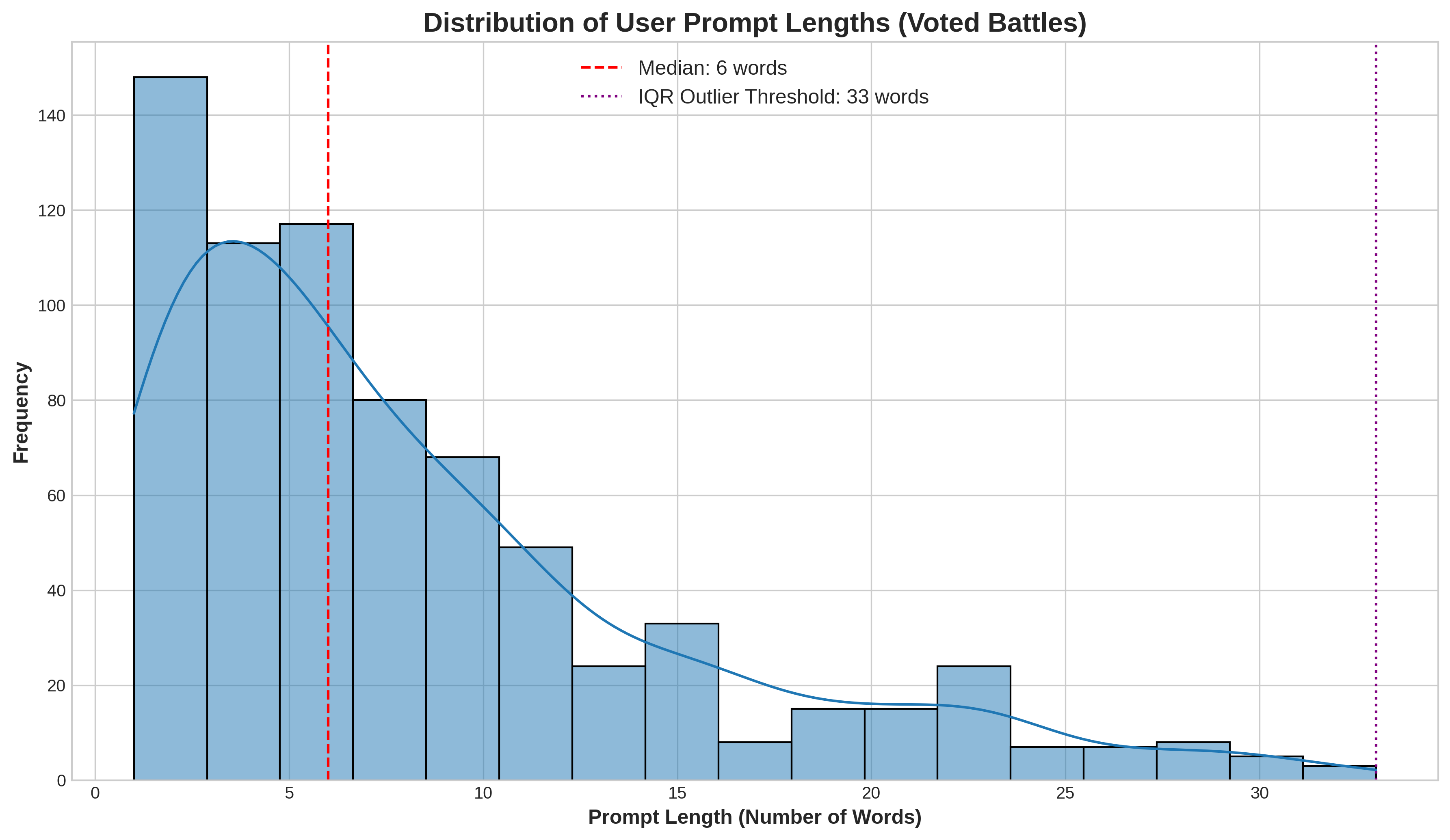} 
    \caption{Distribution of user prompt lengths from voted battles (after removing outliers)} 
    \label{fig:prompt-length-dist} 
\end{figure}

Analyzing the 804 user-written prompts from valid, voted-on battles reveals they are typically concise. 
As shown in \Cref{tab:prompt-stats}, the raw data shows a median prompt length of 7 words, but the mean (18.68) is heavily skewed by a long tail of very descriptive prompts (max 1000 words). 
To get a more accurate picture of typical behavior, we removed 82 extreme outliers using the IQR method (threshold at 33 words). 
\Cref{tab:prompt-stats} shows that the statistics for the remaining 722 prompts are much more focused, with a median length of 6 words and a mean of 8.27.
\Cref{fig:prompt-length-dist} visualizes this post-filtered distribution, confirming that the typical Music Arena user prefers to express their creative ideas in a few words.

\begin{table}[htbp]
\centering
\caption{Descriptive statistics for user prompt lengths (in words), before and after Interquartile Range (IQR) outlier removal.}
\label{tab:prompt-stats}
\begin{tabular}{lcc}
\hline
\textbf{Metric} & \textbf{Raw Data (804 prompts)} & \textbf{Post-IQR (722 prompts)} \\
\hline
count & 804 & 722 \\
mean & 18.68 & 8.27 \\
std & 54.82 & 6.87 \\
min & 1 & 1 \\
50\% (median) & 7 & 6 \\
max & 1000 & 33 \\
\hline
\end{tabular}
\end{table}

\subsection{Commonly Requested Musical Concepts} 
\begin{figure}[htbp]
    \centering  \includegraphics[width=0.8\linewidth]{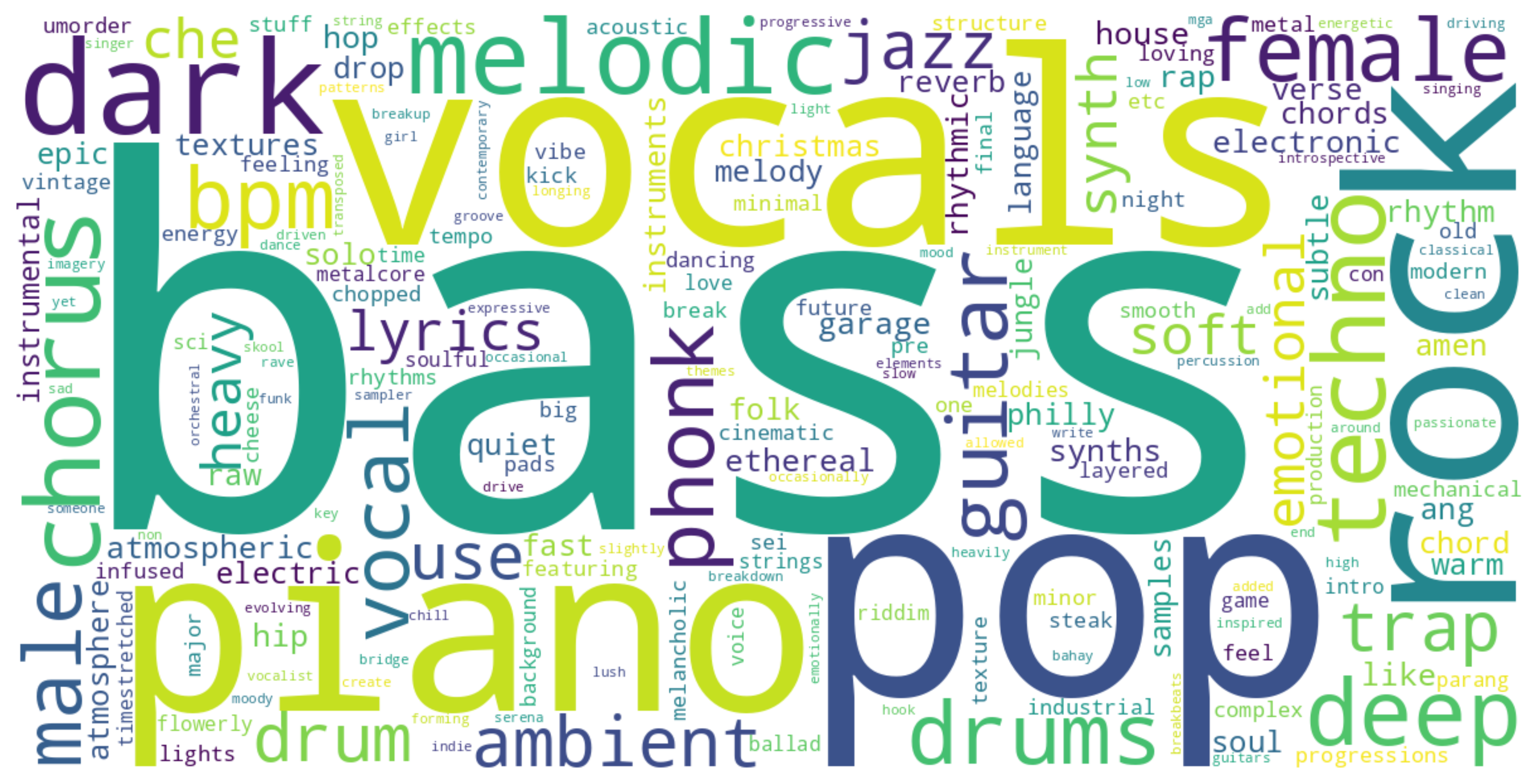} 
    \caption{A word cloud of the most frequent keywords in user prompts from voted battles.}
    \label{fig:word-cloud}
\end{figure}

By analyzing the keywords in 804 user-written prompts, we identified the most common musical genres, instruments, and moods requested by users. \Cref{tab:keywords} lists the most frequent keywords, and Figure \ref{fig:word-cloud} visualizes their prominence.

\begin{table}[htbp]
\centering
\caption{Most frequent keywords in user-written prompts.}
\label{tab:keywords}
\begin{tabular}{lc}
\hline
\textbf{Keyword} & \textbf{Frequency} \\ \hline
bass & 101 \\
pop & 98 \\
vocals & 81 \\
piano & 70 \\
rock & 69 \\
dark & 66 \\
melodic & 66 \\
chorus & 65 \\ \hline
\end{tabular}
\end{table}

\section{Complete example of a Music Arena battle}
\label{sec:battle_example}

\begin{figure}[t]
    \centering
    \includegraphics[width=\textwidth]{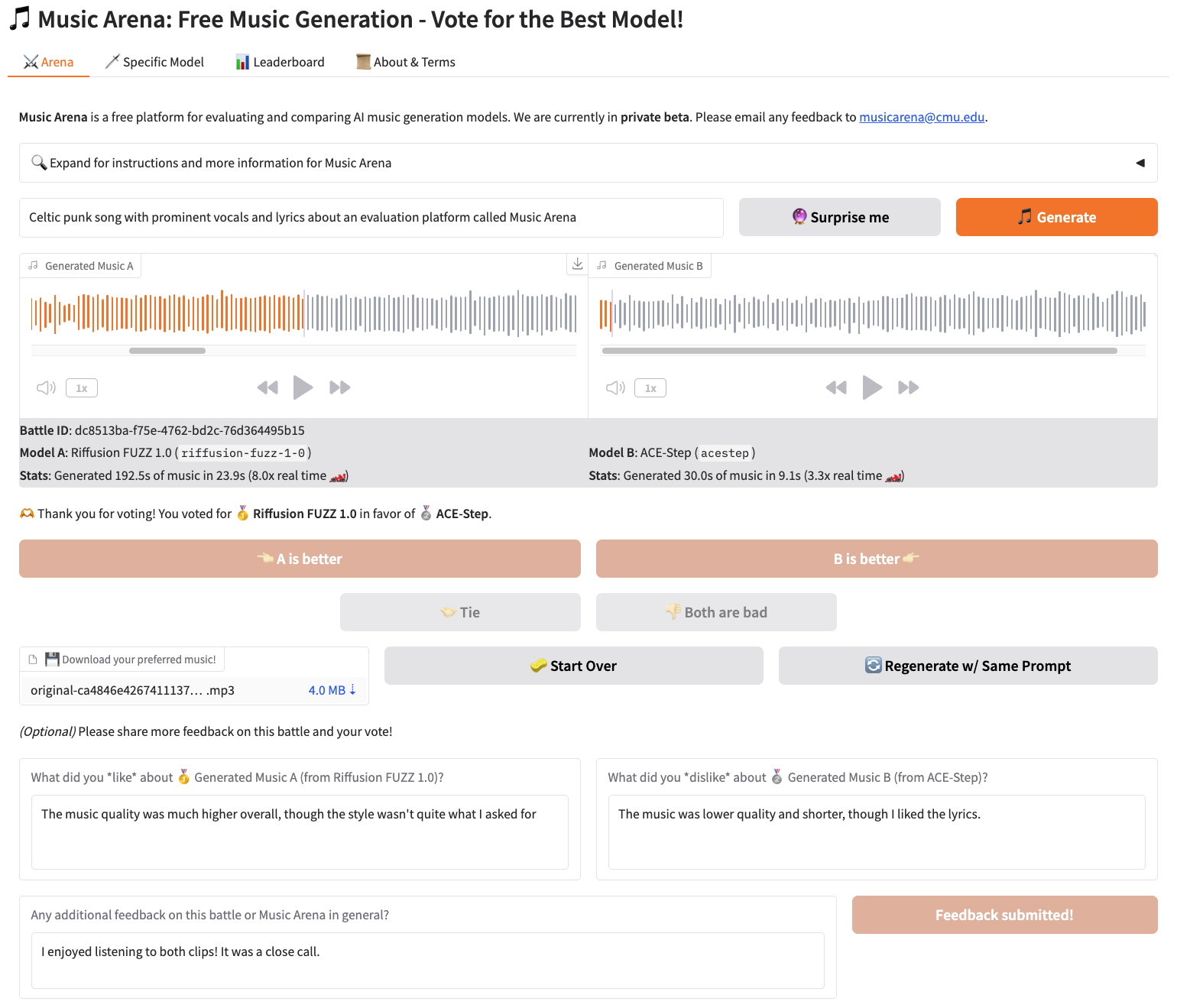}
    \caption{An example of a completed user battle in the Music Arena frontend.}
    \label{fig:battle}
\end{figure}

In~\Cref{fig:battle} we show an example of a completed battle in Music Arena. 
Below we include the complete JSON log from our platform for that same battle, 
highlighting the \detailed{} preference information that we collect.
The corresponding audio is here: \url{https://drive.google.com/drive/folders/1UlueXyaTmef2qw5zwdctXNVgKP9bFa1R?usp=sharing}

\begin{verbatim}
{
  "uuid": "dc8513ba-f75e-4762-bd2c-76d364495b15",
  "gateway_git_hash": "4ae486f55970ce64dad735027f9a8c453d63a6d3:dirty",
  "prompt": {
    "prompt": "Celtic punk song with prominent vocals and lyrics about an evaluation
platform called Music Arena"
  },
  "prompt_detailed": {
    "overall_prompt": "Celtic punk song with prominent vocals and lyrics about an
evaluation platform called Music Arena",
    "instrumental": false,
    "lyrics": null,
    "duration": null
  },
  "prompt_user": {
    "ip": null,
    "salted_ip": "d15300d2f8f7a122a14793494c85057d",
    "fingerprint": null,
    "salted_fingerprint": null
  },
  "prompt_session": {
    "uuid": "42a03157-e3dc-4f00-8a59-1cdc2c221527",
    "create_time": 1753572627.3779469,
    "frontend_git_hash": "4138a182e618f2e7687e4d34bf039cff42275f1e:dirty",
    "ack_tos": "c81b3d54ff3f196eaee354e5317dc6e7",
    "new_battle_times": [
      1753572628.2408764,
      1753572653.6583097
    ]
  },
  "prompt_prebaked": false,
  "prompt_routed": true,
  "a_audio_url": "...",
  "a_metadata": {
    "system_key": {
      "system_tag": "riffusion-fuzz-1-0",
      "variant_tag": "initial"
    },
    "system_git_hash": "1952210249ad28dad600013ccf0ce13130165ca5:dirty",
    "system_time_queued": 1753572662.6410472,
    "system_time_started": 1753572662.6510143,
    "system_time_completed": 1753572686.5897048,
    "gateway_time_started": 1753572661.6347294,
    "gateway_time_completed": 1753572696.3290386,
    "gateway_num_retries": 0,
    "size_bytes": 4167243,
    "lyrics": "[Verse 1]\nAnother blackboard on the wall, spitting names and all,
(la da da)\nI lace my boots but no one calls, just numbers falling,
crawling\nStale beer breath and hate mail\u2014louder than the praise
fails\nScored by some stranger\u2019s scale, (oh oh) chalk dust in the
veins\n[Pre-Chorus]\nWe shout and shake, never for their grades\u2014\nRunning
ragged, never break, chasing after the parade\n[Chorus]\nThis is the Music
Arena\u2014don\u2019t care if you rate us!\nWe bleed out our voices, tearing the
silence\nRaise up our riot, sing for the lost ones\nWhat\u2019s your arena? Ours
is defiance! (Oi! Oi!)\n[Verse 2]\nFour strings snapping, critics clapping
(ha!)\nNotes like knuckles, fists overlapping\nPushed through the carnage,
nobody clean\nYet we shout \u201cViva la m\u00fasica!\u201d\u2014lightning
in-between\n[Breakdown]\n(La la la, la la, hey!)\nNames on the scoreboard,
erased quick as drawn\nWe\u2019re not your numbers\u2014we live on and on!\n(Oh
oh oh, let the echoes run)\n[Chorus]\nThis is the Music Arena\u2014don\u2019t
care if you rate us!\nWe bleed out our voices, tearing the silence\nRaise up our
riot, sing for the lost ones\nWhat\u2019s your arena? Ours is defiance! (Oi!
Oi!)\n[Bridge]\nEntre gritos y cerveza, saltamos sin pena\nTu marca no pesa, en
mi condena\n(Whoa-oh, whoa-oh, vamos!)\n[Solo]\n[Electric guitar and tin
whistle]\n[Chorus]\nThis is the Music Arena\u2014don\u2019t care if you rate
us!\nWe bleed out our voices, tearing the silence\nRaise up our riot, sing for
the lost ones\nWhat\u2019s your arena? Ours is defiance! (Oi! Oi!)\n[Outro]\n(La
la la la, la la la)\nErase us, replace us\u2014we\u2019re loud \u2018til
we\u2019re gone.",
    "sample_rate": 44100,
    "num_channels": 2,
    "duration": 192.496327,
    "checksum": "0008ee5d14cc4bafb3f8ec2fa26f4784"
  },
  "b_audio_url": "...",
  "b_metadata": {
    "system_key": {
      "system_tag": "acestep",
      "variant_tag": "initial"
    },
    "system_git_hash": "1952210249ad28dad600013ccf0ce13130165ca5:dirty",
    "system_time_queued": 1753572655.0246835,
    "system_time_started": 1753572660.6987517,
    "system_time_completed": 1753572669.7986672,
    "gateway_time_started": 1753572655.0225708,
    "gateway_time_completed": 1753572671.424467,
    "gateway_num_retries": 0,
    "size_bytes": 735168,
    "lyrics": "Welcome to the Music Arena, where the notes collide,  \nWith the
bagpipes wailin' and the fiddles in stride,  \nThere are tunes to judge and
rhythms to compare,  \nIn this grand ol' place where musics declare.  \n\nRaise
a pint in the bar where the voices unite,  \nFor in Music Arena, we measure the
heights,  \nOf melodies and harmonies, bold and true,  \nBringing joy to the
hearts of both me and you.  \n\nJoin the chorus loud, in the thunder and the
cheers,  \nMusic Arena's call will ring through the years.  \nA place where the
music is set free,  \nIn the heart of it all, where we want to be.  ",
    "sample_rate": 48000,
    "num_channels": 2,
    "duration": 29.952,
    "checksum": "24b1af8031278bb85e126f9e0bb11028"
  },
  "vote": {
    "a_listen_data": [
      [
        "PLAY",
        1753572708.6986423
      ],
      [
        "TICK",
        1753572709.9190872
      ],
      ...,
      [
        "TICK",
        1753572729.919615
      ],
      [
        "PAUSE",
        1753572731.188438
      ],
      [
        "TICK",
        1753572731.2903912
      ],
      ...,
      [
        "TICK",
        1753572736.7407818
      ],
      [
        "PLAY",
        1753572763.5559134
      ],
      [
        "PAUSE",
        1753572789.6293015
      ]
    ],
    "b_listen_data": [
      [
        "PLAY",
        1753572731.9962952
      ],
      [
        "TICK",
        1753572733.203671
      ],
      ...,
      [
        "TICK",
        1753572736.2387252
      ],
      [
        "PAUSE",
        1753572761.9931803
      ],
      [
        "PLAY",
        1753572762.144039
      ],
      [
        "PAUSE",
        1753572762.799093
      ]
    ],
    "preference": "A",
    "preference_time": 1753572791.0873723,
    "feedback": "I enjoyed listening to both clips! It was a close call.",
    "a_feedback": "The music quality was much higher overall, though the style
wasn't quite what I asked for",
    "b_feedback": "The music was lower quality and shorter, though I liked the
lyrics.",
    "feedback_time": 1753572842.6993084
  },
  "vote_user": {
    "ip": null,
    "salted_ip": "d15300d2f8f7a122a14793494c85057d",
    "fingerprint": null,
    "salted_fingerprint": null
  },
  "vote_session": {
    "uuid": "42a03157-e3dc-4f00-8a59-1cdc2c221527",
    "create_time": 1753572627.3779469,
    "frontend_git_hash": "4138a182e618f2e7687e4d34bf039cff42275f1e:dirty",
    "ack_tos": "c81b3d54ff3f196eaee354e5317dc6e7",
    "new_battle_times": [
      1753572628.2408764,
      1753572653.6583097
    ]
  },
  "timings": [
    [
      "parse",
      1753572653.815334
    ],
    [
      "generate",
      1753572653.815521
    ],
    [
      "route",
      1753572653.815524
    ],
    [
      "sample_pair",
      1753572655.0172503
    ],
    [
      "generate_parallel_start",
      1753572655.0173497
    ],
    [
      "health_check_riffusion-fuzz-1-0:initial_start",
      1753572655.0174298
    ],
    [
      "health_check_acestep:initial_start",
      1753572655.0176365
    ],
    [
      "health_check_acestep:initial_end",
      1753572655.0225341
    ],
    [
      "generate_acestep:initial_start",
      1753572655.0225701
    ],
    [
      "health_check_riffusion-fuzz-1-0:initial_end",
      1753572661.634688
    ],
    [
      "generate_riffusion-fuzz-1-0:initial_start",
      1753572661.6347291
    ],
    [
      "generate_acestep:initial_end",
      1753572671.4905503
    ],
    [
      "generate_riffusion-fuzz-1-0:initial_end",
      1753572696.4137614
    ],
    [
      "generate_parallel_end",
      1753572696.4139218
    ],
    [
      "create_battle_obj",
      1753572696.4139223
    ],
    [
      "upload_audio",
      1753572696.415068
    ],
    [
      "upload_metadata",
      1753572697.0474696
    ],
    [
      "vote",
      1753572791.2476099
    ],
    [
      "vote",
      1753572842.7061708
    ]
  ]
}
\end{verbatim}

\end{document}